\documentclass[12pt]{article}
\usepackage{graphicx}
\usepackage{amsmath}
\usepackage{amssymb}
\usepackage{caption2}
\begin{document}
\bibliographystyle{prsty}
\begin{center}
{\large {\bf \sc{   Strong decays $\Sigma^{*} \to
\Sigma\pi,\Lambda \pi$ and related strong coupling constant  }}} \\[2mm]
Zhi-Gang Wang \footnote{E-mail:wangzgyiti@yahoo.com.cn.  }   \\
  Department of Physics, North China Electric Power University, Baoding 071003, P. R. China
 \end{center}

\begin{abstract}
In this article, we calculate the strong coupling constant $g$ among
the decuplet baryons, the octet baryons and the pseudoscalar mesons
in the heavy baryon chiral perturbation theory with the light-cone
QCD sum rules, and study the strong decays $\Sigma^* \to \Lambda
\pi,\Sigma \pi$. The numerical value of the strong coupling constant
$g$ is consistent with our previous calculation, the central values
lead to small $SU(3)$ breaking effects, less than $6\%$; and no
definitive  conclusion can be drawn due to the large uncertainties.
\end{abstract}


\section{Introduction}

In the (heavy) baryon chiral perturbation theory, the resonant
baryon states   are usually assumed  to be  very heavy  and
integrated out,  their effects are represented  by a finite piece of
counterterms. The  mass difference between the decuplet baryons and
the octet baryons is rather small, about $300\, \rm{MeV}$,  and the
coupling constant among the decuplet baryons, the octet baryons and
the pseudoscalar mesons is rather large \cite{PDG}. For example, the
$\Delta(1232)$ resonance dominates many nuclear phenomena at
energies above the pion-production threshold. It is almost an ideal
elastic $\pi N$ resonance, and decays into the nucleon  and pion
($\Delta \to N\pi$) with the branching  fraction about $99\%$
\cite{PDG}. It is useful to include the decuplet baryons as an
explicit degree of freedom in the effective lagrangian.   In the
small scale expansion approach (which builds upon the  heavy baryon
chiral perturbation theory), the
 nucleon and $\Delta$ degrees of freedom are treated
simultaneously \cite{HBCHPT}.

The phenomenological chiral lagrangian can be written as
\begin{equation}
\mathcal{ L}=-{\cal C}\left[\overline{T}^\mu u_\mu
B+\overline{B}u_\mu T^\mu \right] \, ,
\end{equation}
where
\begin{eqnarray}
u_\mu &=&{i\over 2} \{\xi^\dagger, \partial_\mu \xi\} \, , \nonumber\\
 \xi &=& \exp(
\frac{i\phi}{f_\pi})\, , \nonumber\\
\phi&=&\left(
\begin{array}{ccc}
\frac{\pi^0}{\sqrt2}+\frac{\eta}{\sqrt6}&\pi^+&K^+\\
\pi^-&-\frac{\pi^0}{\sqrt2}+\frac{\eta}{\sqrt6}&K^0\\
K^-&\overline{K}^0&-\frac{2}{\sqrt6}\eta
\end{array}\right) \, ,\nonumber\\
B&=&\left(
\begin{array}{ccc}
\frac{\Sigma^0}{\sqrt2}+\frac{\Lambda}{\sqrt6}&\Sigma^+&p\\
\Sigma^-&-\frac{\Sigma^0}{\sqrt2}+\frac{\Lambda}{\sqrt6}&n\\
\Xi^-&\Xi^0&-\frac{2}{\sqrt6}\Lambda
\end{array}\right) \, ,
\end{eqnarray}
and
\begin{eqnarray}
T_{uuu}=\Delta^{++}, \qquad T_{uud}=\frac{\Delta^{+}}{\sqrt3}, \qquad T_{udd}=\frac{\Delta^{0}}{\sqrt3}, \qquad T_{ddd}=\Delta^{-}, \qquad T_{uus}=\frac{\Sigma^{*+}}{\sqrt3},\nonumber\\
T_{uds}=\frac{\Sigma^{*0}}{\sqrt6}, \qquad
T_{dds}=\frac{\Sigma^{*-}}{\sqrt3}, \qquad
T_{uss}=\frac{\Xi^{*0}}{\sqrt3}, \qquad
T_{dss}=\frac{\Xi^{*-}}{\sqrt3}, \qquad T_{sss}=\Omega^- \, ,
\end{eqnarray}
 the $T_\mu$ are the Rarita-Schwinger fields of the decuplet baryons $T_i$,  the $f_\pi$ is the decay
 constant of the $\pi$. From the
chiral lagrangian $\mathcal{L}$, we can obtain the following
relations,
\begin{eqnarray}
g_{\Delta^{++} p\pi^+}&=&\frac{\mathcal{C}}{f_\pi}\, ,\nonumber \\
g_{\Sigma^{*+} \Sigma^0 \pi^+ } &=&-\frac{\mathcal{C}}{\sqrt{6}f_\pi} \, , \nonumber \\
g_{\Sigma^{*+} \Lambda \pi^+ } &=&-\frac{\mathcal{C}}{\sqrt{2}f_\pi}
\, ,
\end{eqnarray}
the  coupling constant  $\mathcal{C}$ is a basic parameter,   which
can be fitted phenomenologically or calculated with some theoretical
approaches, we introduce a parameter $g$ with
$g=\frac{\mathcal{C}}{f_\pi}$ to simplify the notation.  We
terminate the Taylor series $\xi = \exp(
\frac{i\phi}{f_\pi})=1+\frac{i\phi}{f_\pi}+\frac{1}{2!}\left(\frac{i\phi}{f_\pi}\right)^2+\cdots$
at leading order $\mathcal {O}(\phi)$ and approximate $u_\mu
={i\over 2} \{\xi^\dagger,
\partial_\mu \xi\}\approx -\frac{\partial_\mu\phi}{f_\pi}$ to obtain
the three relations in Eq.(4).  The higher order terms of the Taylor
series $\xi =
 1+\frac{i\phi}{f_\pi}+\frac{1}{2!}\left(\frac{i\phi}{f_\pi}\right)^2+\cdots$
 have contributions to the strong coupling constants $g_{\Delta^{++}
 p\pi^+}$, $g_{\Sigma^{*+} \Sigma^0 \pi^+ }$ and $g_{\Sigma^{*+} \Lambda \pi^+
 }$,  their  effects can be taken into account   with the replacement
 $\frac{\mathcal{C}}{f_\pi}\rightarrow \frac{\mathcal{C}}{f_\pi}
 \left(1+\frac{\alpha}{f_\pi}+\frac{\beta}{f_\pi^2}+\cdots\right)$,
 where the coefficients $\alpha$ and $\beta$ originate from the corresponding chiral
 loops. In this article, we take the leading order approximation.

 Thereafter we will introduce the notations $g_N$, $g_\Sigma$
and $g_\Lambda$ to represent the strong coupling constant $g$ from
the $g_{\Delta^{++} p\pi^+}$, $g_{\Sigma^{*+} \Sigma^0 \pi^+ }$ and
$g_{\Sigma^{*+} \Lambda \pi^+ }$ respectively, and study the strong
coupling constants $g_\Lambda$, $g_\Sigma$ and the $SU(3)$ breaking
effects with the light-cone QCD sum rules.

From the Particle Data Group \cite{PDG}, we can see that the
following strong decays are kinematically allowed,
\begin{eqnarray}
\Delta &\to& p \pi \, , \nonumber\\
\Sigma^{*} &\to& \Sigma\pi,\Lambda \pi\,  ,\nonumber \\
\Xi^* &\to& \Xi \pi \, ,
\end{eqnarray}
the strong decays $\Sigma^{*} \to \Sigma\pi,\Lambda \pi$ are ideal
channels to study the $SU(3)$ breaking effects as the constituent
quark contents of the baryons $\Sigma^*$, $\Sigma$ and $\Lambda$ are
 $uds$ or $uus$.

 In a previous work, we have calculated the strong coupling constant $g_{N}$
  with the light-cone QCD sum rules, and studied  the decay width $\Gamma_{\Delta \to p\pi}$ \cite{Wang0707}. The strong
 coupling constants among the octet baryons, the vector and pseudoscalar mesons
 $g_{ NNV}$ and $g_{NNP}$ have been calculated with the
 light-cone QCD sum rules \cite{Baryon1,Baryon2,Baryon3,Baryon4}.

 The light-cone QCD sum
rules  carry out the operator product expansion near the light-cone
$x^2\approx 0$ instead of the short distance $x\approx 0$ while the
nonperturbative hadronic matrix elements  are parameterized by the
light-cone distribution amplitudes   instead of  the vacuum
condensates \cite{LCSR1,LCSR2,LCSR3,LCSR4,LCSR5,LCSRreview}. The
nonperturbative
 parameters in the light-cone distribution amplitudes are calculated with  the conventional QCD  sum rules
 and the  values are universal \cite{SVZ79,PRT85}.

The article is arranged as: in Section 2, we derive the strong
coupling constants  $g_{\Lambda}$ and $g_{\Sigma}$ with the
light-cone QCD sum rules; in Section 3, the numerical result and
discussion; and Section 4 is reserved for conclusion.

\section{Strong coupling constants   $g_{\Lambda}$ and $g_{\Sigma}$ with light-cone QCD sum rules}

In the following, we write down the
 two-point correlation functions $\Pi^{\Lambda/\Sigma}_\mu(p,q)$,
\begin{eqnarray}
\Pi^{\Lambda/\Sigma}_\mu(p,q)&=&i \int d^4x \, e^{-i q \cdot x} \,
\langle 0 |T\left\{ J_{\Lambda/\Sigma}(0)\bar{J}_\mu(x)\right\}|\pi(p)\rangle \, , \\
J_\Lambda(x)&=& \sqrt{\frac{2}{3}}\epsilon_{abc} \left[ u^T_a(x)C\gamma_\mu s_b(x) \gamma_5 \gamma^\mu d_c(x)-d^T_a(x)C\gamma_\mu s_b(x) \gamma_5 \gamma^\mu u_c(x) \right]\, ,  \nonumber \\
J_\Sigma(x)&=& \sqrt{2}\epsilon_{abc} \left[ u^T_a(x)C\gamma_\mu s_b(x) \gamma_5 \gamma^\mu d_c(x)+d^T_a(x)C\gamma_\mu s_b(x) \gamma_5 \gamma^\mu u_c(x) \right]\, ,  \nonumber \\
J_\mu(x)&=& \frac{\epsilon_{abc}}{\sqrt{3}}\left[
2u^T_a(x)C\gamma_\mu s_b(x) u_c(x)+u^T_a(x)C\gamma_\mu u_b(x) s_c(x)
\right]\, ,
\end{eqnarray}
where the baryon currents $J_\Sigma(x)$, $J_\Lambda(x)$ and
$J_\mu(x)$
 interpolate the octet baryons $\Sigma$, $\Lambda$ and the  decuplet baryon
 $\Sigma^*$,  respectively \cite{Ioffe1,Ioffe2,Ioffe3,Ioffe4}, the external  state $\pi$ has the
four momentum $p_\mu$ with $p^2=m_\pi^2$ .  The correlation
functions $\Pi_{\mu}(p,q)$ (sometime we will smear the indexes
$\Lambda$ and $\Sigma$ for simplicity) can be decomposed as
\begin{eqnarray}
\Pi_{\mu }(p,q)&=&\Pi\sigma_{\alpha\beta}p^\alpha q^\beta
p_\mu+\Pi_{A1}p_\mu+\Pi_{A2}\not\!\!q p_\mu+\Pi_{A3}\not\!\!pp_\mu+
\nonumber\\
&&\Pi_{B1}q_\mu+\Pi_{B2}\not\!\!q
q_\mu+\Pi_{B3}\not\!\!pq_\mu+\Pi_{B4}\sigma_{\alpha\beta}p^\alpha
q^\beta
q_\mu+\nonumber\\
&&\Pi_{C1}\gamma_\mu+\Pi_{C2}\not\!\!q\gamma_\mu+\Pi_{C3}\not\!\!p\gamma_\mu+\Pi_{C4}\epsilon_{\mu\nu\alpha\beta}\gamma^\nu
\gamma_5 p^\alpha q^\beta
\end{eqnarray}
due to the Lorentz invariance, where the $\Pi$ and $\Pi_i$ are
Lorentz invariant functions of  $p$ and $q$. In this article, we
choose the tensor structure $\sigma_{\alpha\beta}p^\alpha q^\beta
p_\mu$ for analysis.

Basing on the quark-hadron duality \cite{SVZ79,PRT85}, we can insert
a complete set  of intermediate hadronic states with the same
quantum numbers as the current operators $J_{\Lambda/\Sigma}(x)$ and
$J_\mu(x)$ into the correlation functions $\Pi_{\mu}(p,q)$  to
obtain the hadronic representation. After isolating the ground state
contributions from the pole terms of the baryons $\Lambda/\Sigma$
and $\Sigma^*$, we get the following results,
\begin{eqnarray}
\Pi^\Lambda_{\mu }(p,q)&=&\frac{\langle0| J_\Lambda(0)|
\Lambda(q+p)\rangle\langle \Lambda(q+p)| \Sigma^*(q) \pi(p) \rangle
\langle \Sigma^*(q)|\bar{J}_\mu(0)| 0\rangle}
  {\left\{M_{\Lambda}^2-(q+p)^2\right\}\left\{M_{\Sigma^*}^2-q^2\right\}}  + \cdots \nonumber \\
&=& \frac{\lambda_\Lambda
\lambda_{\Sigma^*}}{\left\{M_\Lambda^2-(q+p)^2\right\}\left\{M_{\Sigma^*}^2-q^2\right\}}
\left\{ \frac{g_{\Sigma^* \Lambda
\pi}}{3}\sigma_{\alpha\beta}p_\alpha q_\beta
p_\mu+\cdots\right\}+\cdots \, ,\\
\Pi^\Sigma_{\mu }(p,q)&=&\frac{\langle0| J_\Sigma(0)|
\Sigma(q+p)\rangle\langle \Sigma(q+p)| \Sigma^*(q) \pi(p) \rangle
\langle \Sigma^*(q)|\bar{J}_\mu(0)| 0\rangle}
  {\left\{M_{\Sigma}^2-(q+p)^2\right\}\left\{M_{\Sigma^*}^2-q^2\right\}}  + \cdots \nonumber \\
&=& \frac{\lambda_\Sigma
\lambda_{\Sigma^*}}{\left\{M_\Sigma^2-(q+p)^2\right\}\left\{M_{\Sigma^*}^2-q^2\right\}}
\left\{ \frac{g_{\Sigma^* \Sigma
\pi}}{3}\sigma_{\alpha\beta}p_\alpha q_\beta
p_\mu+\cdots\right\}+\cdots \, ,
\end{eqnarray}
where the following definitions have been used,
\begin{eqnarray}
\langle 0| J_{\Lambda/\Sigma} (0)|\Lambda/\Sigma(p)\rangle &=&\lambda_{\Lambda/\Sigma} U(p,s) \, , \nonumber \\
\langle 0| J_\mu (0)|\Sigma^*(p)\rangle &=&\lambda_{\Sigma^*} U_\mu(p,s) \, , \nonumber \\
\sum_sU(p,s)\overline {U}(p,s)&=&\!\not\!{p}+M_{\Lambda/\Sigma} \, , \nonumber \\
\sum_s U_\mu(p,s) \overline{U}_\nu(p,s)
&=&-(\!\not\!{p}+M_{\Sigma^*})\left\{ g_{\mu\nu}-\frac{\gamma_\mu
\gamma_\nu}{3}-\frac{2p_\mu p_\nu}{3M_{\Sigma^*}^2}+\frac{p_\mu
\gamma_\nu-p_\nu \gamma_\mu}{3M_{\Sigma^*}} \right\} \, , \nonumber\\
\langle \Lambda/\Sigma(q')| \Sigma^*(q) \pi(p)
\rangle&=&ig_{\Sigma^* \Lambda/\Sigma
\pi}\overline{U}(q',s')U_\mu(q,s)p^\mu\, .
\end{eqnarray}

 The current $J_\mu(x)$ couples
not only to the isospin $I=\frac{3}{2}$ and spin-parity
$J^P=\frac{3}{2}^+$ states, but also to the isospin $I=\frac{3}{2}$
and spin-parity $J^P=\frac{1}{2}^-$ states.
 For a generic $\frac{1}{2}^-$ resonance   $\widetilde{\Sigma^*}$ \cite{Braun06},
\begin{eqnarray}
 \langle0|J_{\mu}(0)|\widetilde{\Sigma^*}(p)\rangle=\lambda_{*}
 (\gamma_{\mu}-4\frac{p_{\mu}}{M_{*}})U^{*}(p,s) \, ,
\end{eqnarray}
where $\lambda^{*}$ is  the  pole residue  and $M_{*}$ is the mass.
The spinor $U^*(p,s)$  satisfies the usual Dirac equation
$(\not\!\!p-M_{*})U^{*}(p)=0$. If we take the phenomenological
lagrangian,
\begin{eqnarray}
\mathcal {L}(x)&=& g_{\widetilde{\Sigma^*} \Lambda/\Sigma
\pi}\left\{\overline{\widetilde{\Sigma^*}}(x) \Lambda/\Sigma(x)
\pi(x) +\overline{\Lambda}/\overline{\Sigma}(x)
\widetilde{\Sigma^*}(x) \pi(x) \right\} \, ,
\end{eqnarray}
which corresponds to $\langle \Lambda/\Sigma(q')|
\widetilde{\Sigma^*}(q) \pi(p) \rangle=g_{\widetilde{\Sigma^*}
\Lambda/\Sigma \pi}\overline{U}(q',s')U^*(q,s)$, the contributions
from the $\frac{1}{2}^-$ states can be written as
\begin{eqnarray}
\Pi^{\Lambda/\Sigma}_\mu(p,q)&=&\frac{g_{\widetilde{\Sigma^*}
\Lambda/\Sigma \pi}\lambda_{\Lambda/\Sigma}
\lambda_*}{\left\{M_{\Lambda/\Sigma}^2-(q+p)^2\right\}\left\{M_*^2-q^2\right\}}
\left\{ (\not\!\!p+\not\!\!q +M_{\Lambda/\Sigma})(\not\!\!q +M_*)(\gamma_{\mu}-4\frac{q_{\mu}}{M_{*}})\right\}\nonumber\\
&&+\cdots \nonumber\\
&=&\Pi_{D}\not\!\!q p_\mu +\Pi_{E1}q_\mu+\Pi_{E2}\not\!\!q
q_\mu+\Pi_{E3}\not\!\!pq_\mu+\Pi_{E4}\sigma_{\alpha\beta}p^\alpha
q^\beta
q_\mu+\nonumber\\
&&\Pi_{F1}\gamma_\mu+\Pi_{F2}\not\!\!q\gamma_\mu+\Pi_{F3}\not\!\!p\gamma_\mu+\Pi_{F4}\epsilon_{\mu\nu\alpha\beta}\gamma^\nu
\gamma_5 p^\alpha q^\beta \, ,
\end{eqnarray}
where the  $\Pi_i$ are Lorentz invariant functions of  $p$ and $q$.
 If we choose the tensor structure $\sigma_{\alpha\beta}p_\alpha q_\beta
p_\mu$, the $\widetilde{\Sigma^*}$  has no contaminations.

In the following, we briefly outline the  operator product expansion
for the correlation functions  $\Pi_{\mu }(p,q)$  in perturbative
QCD theory. The calculations are performed at the large space-like
momentum regions $(q+p)^2\ll 0$  and $q^2\ll 0$, which correspond to
the small light-cone distance $x^2\approx 0$ required by the
validity of the operator product expansion approach. We write down
the "full" propagator of a massive light  quark in the presence of
the quark and gluon condensates firstly \cite{LCSR1,PRT85},
\begin{eqnarray}
\langle0| T[q_a(x)q_b(0)]|0\rangle &=&
\frac{i\delta_{ab}\!\not\!{x}}{ 2\pi^2x^4}
-\frac{\delta_{ab}m_q}{4\pi^2x^2}-\frac{\delta_{ab}}{12}\langle
\bar{q}q\rangle +\frac{i\delta_{ab}}{48}m_q
\langle\bar{q}q\rangle-\frac{\delta_{ab}x^2}{192}\langle
\bar{q}g_s\sigma Gq\rangle
\nonumber\\
&& +\frac{i\delta_{ab}x^2}{1152}m_q\langle \bar{q}g_s\sigma
Gq\rangle \!\not\!{x}\nonumber\\
&&-\frac{i}{16\pi^2x^2} \int_0^1 dv
\left[(1-v)g_sG_{\mu\nu}(vx)\!\not\!{x}
\sigma^{\mu\nu}+vg_sG_{\mu\nu}(vx)\sigma^{\mu\nu}
\!\not\!{x}\right]  \nonumber\\
&&+\cdots \, ,
\end{eqnarray}
then contract the quark fields in the correlation functions
$\Pi_\mu(p,q)$ with the Wick theorem, and obtain the following
results,
\begin{eqnarray}
\Pi^\Lambda_\mu(p,q)&=&\frac{2\sqrt{2}}{3}i\epsilon_{abc}\epsilon_{a'b'c'}
\int d^4x e^{-i q
\cdot x} \nonumber \\
&&\left\{ Tr\left[ \gamma_\mu  CS_{bb'}^T(-x)C\gamma_\alpha U_{aa'}(-x)\right]\gamma_5\gamma^\mu \langle 0|d_c(0)\bar{u}_{c'}(x) |\pi(p)\rangle\right.\nonumber \\
&&-Tr\left[ \gamma_\mu  CS_{bb'}^T(-x)C\gamma_\alpha \langle 0|d_a(0)\bar{u}_{a'}(x) |\pi(p)\rangle\right]\gamma_5\gamma^\mu U_{cc'}(-x) \nonumber \\
&&-\gamma_5 \gamma^\alpha \langle
0|d_c(0)\bar{u}_{a'}(x)|\pi(p)\rangle \gamma_\mu C S^T_{bb'}(-x)C
\gamma_\alpha U_{ac'}(-x) \nonumber \\
&&+\gamma_5 \gamma^\alpha U_{ca'}(-x) \gamma_\mu C S^T_{bb'}(-x)C
\gamma_\alpha \langle
0|d_a(0)\bar{u}_{c'}(x)|\pi(p)\rangle \nonumber \\
&&-\gamma_5 \gamma^\alpha \langle
0|d_c(0)\bar{u}_{b'}(x)|\pi(p)\rangle \gamma_\mu C U^T_{aa'}(-x)C
\gamma_\alpha S_{bc'}(-x) \nonumber \\
 &&\left. +\gamma_5 \gamma^\alpha U_{cb'}(-x) \gamma_\mu C \left[\langle
0|d_a(0)\bar{u}_{a'}(x)|\pi(p)\rangle\right]^T C \gamma_\alpha
S_{bc'}(-x)\right\}\, , \\
\Pi^\Sigma_\mu(p,q)&=&\frac{2\sqrt{2}}{\sqrt{3}}i\epsilon_{abc}\epsilon_{a'b'c'}
\int d^4x e^{-i q
\cdot x} \nonumber \\
&&\left\{ Tr\left[ \gamma_\mu  CS_{bb'}^T(-x)C\gamma_\alpha U_{aa'}(-x)\right]\gamma_5\gamma^\mu \langle 0|d_c(0)\bar{u}_{c'}(x) |\pi(p)\rangle\right.\nonumber \\
&&+Tr\left[ \gamma_\mu  CS_{bb'}^T(-x)C\gamma_\alpha \langle 0|d_a(0)\bar{u}_{a'}(x) |\pi(p)\rangle\right]\gamma_5\gamma^\mu U_{cc'}(-x) \nonumber \\
&&-\gamma_5 \gamma^\alpha \langle
0|d_c(0)\bar{u}_{a'}(x)|\pi(p)\rangle \gamma_\mu C S^T_{bb'}(-x)C
\gamma_\alpha U_{ac'}(-x) \nonumber \\
&&-\gamma_5 \gamma^\alpha U_{ca'}(-x) \gamma_\mu C S^T_{bb'}(-x)C
\gamma_\alpha \langle
0|d_a(0)\bar{u}_{c'}(x)|\pi(p)\rangle \nonumber \\
&&-\gamma_5 \gamma^\alpha \langle
0|d_c(0)\bar{u}_{b'}(x)|\pi(p)\rangle \gamma_\mu C U^T_{aa'}(-x)C
\gamma_\alpha S_{bc'}(-x) \nonumber \\
 &&\left. -\gamma_5 \gamma^\alpha U_{cb'}(-x) \gamma_\mu C \left[\langle
0|d_a(0)\bar{u}_{a'}(x)|\pi(p)\rangle\right]^T C \gamma_\alpha
S_{bc'}(-x)\right\}\, .
\end{eqnarray}
Perform the following Fierz re-ordering to extract the contributions
from the two-particle and three-particle $\pi$-meson light-cone
distribution amplitudes respectively,
\begin{eqnarray}
q^a_\alpha(0) \bar{q}^b_\beta(x)&=&-\frac{1}{12}
\delta_{ab}\delta_{\alpha\beta}\bar{q}(x)q(0)
-\frac{1}{12}\delta_{ab}(\gamma^\mu)_{\alpha\beta}\bar{q}(x)\gamma_\mu
q(0) \nonumber\\
&&-\frac{1}{24}\delta_{ab}(\sigma^{\mu\nu})_{\alpha\beta}\bar{q}(x)\sigma_{\mu\nu}q(0) \nonumber\\
&&+\frac{1}{12}\delta_{ab}(\gamma^\mu
\gamma_5)_{\alpha\beta}\bar{q}(x)\gamma_\mu \gamma_5 q(0)\nonumber\\
&&+\frac{1}{12}\delta_{ab}(i \gamma_5)_{\alpha\beta}\bar{q}(x)i
\gamma_5 q(0) \, , \\
q^a_\alpha(0)
\bar{q}^b_\beta(x)G^{ba}_{\lambda\tau}(vx)&=&-\frac{1}{4}
 \delta_{\alpha\beta}\bar{q}(x)G_{\lambda\tau}(vx)q(0)
-\frac{1}{4} (\gamma^\mu)_{\alpha\beta}\bar{q}(x)\gamma_\mu
G_{\lambda\tau}(vx)
q(0) \nonumber\\
&&-\frac{1}{8} (\sigma^{\mu\nu})_{\alpha\beta}\bar{q}(x)\sigma_{\mu\nu}G_{\lambda\tau}(vx)q(0) \nonumber\\
&&+\frac{1}{4} (\gamma^\mu
\gamma_5)_{\alpha\beta}\bar{q}(x)\gamma_\mu \gamma_5 G_{\lambda\tau}(vx)q(0)\nonumber\\
&&+\frac{1}{4} (i \gamma_5)_{\alpha\beta}\bar{q}(x)i
\gamma_5G_{\lambda\tau}(vx) q(0) \, ,
\end{eqnarray}
 and substitute the hadronic matrix elements (such as the $ \langle
0| {\bar u} (x) \gamma_\mu \gamma_5 d(0) |\pi(p)\rangle$,  $ \langle
0| {\bar u} (x)g_s\sigma_{\mu\nu}\gamma_5 G_{\alpha\beta}(vx) d(0)
|\pi(p)\rangle$, $ \langle 0| {\bar u} (x) \sigma_{\mu\nu}\gamma_5
d(0) |\pi(p)\rangle$, etc.)  with
 the corresponding $\pi$-meson light-cone distribution amplitudes, finally we
obtain the spectral densities  at the coordinate space. Once the
spectral densities in the coordinate space are obtained,
 we can translate them   into the
 momentum space with the $D=4+2\epsilon$ dimensional Fourier transform,

\begin{eqnarray}
6\sqrt{2}\Pi_\Lambda&=&-\frac{f_\pi}{\pi^2}
 \int_0^1 du u\phi_{\pi}(u)\frac{\Gamma(\epsilon)}{(-Q^2)^\epsilon}+\frac{3f_\pi m_\pi^2}{4\pi^2}
 \int_0^1 du u A(u)\frac{\Gamma(1)}{(-Q^2)^{1}}
\nonumber \\
&&-4  m_s \langle \bar{s}s\rangle f_\pi
 \int_0^1 du u\phi_{\pi}(u)\frac{\Gamma(2)}{(-Q^2)^{2}}  \nonumber \\
 &&+  \frac{2m_s \langle g_s \sigma G\bar{s}s\rangle f_\pi}{3}
 \int_0^1 du u\phi_{\pi}(u)\frac{\Gamma(3)}{(-Q^2)^{3}}\nonumber \\
 &&- \frac{8 [\langle \bar{q}q\rangle-\langle \bar{s}s\rangle] f_\pi m_\pi^2}{9[m_u+m_d]}
 \int_0^1 du u\phi_{\sigma}(u)\frac{\Gamma(2)}{(-Q^2)^{2}}\nonumber \\
 &&+ \frac{2 [\langle \bar{q}g_s \sigma Gq\rangle-\langle \bar{s}g_s \sigma Gs\rangle] f_\pi m_\pi^2}{9[m_u+m_d]}
 \int_0^1 du u\phi_{\sigma}(u)\frac{\Gamma(3)}{(-Q^2)^{3}}\nonumber \\
 &&- \frac{ m_s f_\pi m_\pi^2}{3\pi^2[m_u+m_d]}
 \int_0^1 du u\phi_{\sigma}(u)\frac{\Gamma(1)}{(-Q^2)^{1}}\nonumber \\
&&-\frac{2f_\pi}{3}\langle\frac{ \alpha_sGG}{\pi}\rangle
 \int_0^1 du u\phi_{\pi}(u)\frac{\Gamma(2)}{(-Q^2)^{2}}\nonumber \\
&& +\frac{f_{3\pi}[4\langle \bar{q}q\rangle+\langle \bar{s}s\rangle]}{12}\int_0^1 dvv \int_0^1 d\alpha_g \int_0^{1-\alpha_g} d\alpha_u\nonumber \\
&&   \frac{\Gamma(2)}{(-Q^2)^{2}}\mid_{u=\alpha_u+v\alpha_g} \phi_{3\pi}(\alpha_u,\alpha_g,1-\alpha_u-\alpha_g)\nonumber \\
&&+\frac{f_\pi m_\pi^2}{4\pi^2} \int_0^1 dv \int_0^1 d\alpha_g \int_0^{1-\alpha_g} d\alpha_u u \frac{\Gamma(1)}{(-Q^2)^{1}}\mid_{u=\alpha_u+v\alpha_g}\nonumber \\
&& \left[-13(1-3v)V_{\perp}+12(1-2v)A_{\parallel}+12(1-v)A_{\perp} \right](\alpha_u,\alpha_g,1-\alpha_u-\alpha_g) \nonumber \\
&&+\frac{6f_\pi m_\pi^2}{\pi^2} \int_0^1 dv \int^1_0 d\alpha_g
\int_0^{1-\alpha_g} d\alpha_u
\int_0^{\alpha_u} d\alpha  \frac{\Gamma( 1)}{(-Q^2)^{ 1}}\mid_{u=\alpha_u+v\alpha_g} \nonumber\\
&&\left[V_{\parallel}+V_{\perp}+(1-2v)(A_{\parallel}+A_{\perp})
\right](\alpha,\alpha_g,1-\alpha-\alpha_g) \nonumber\\
&&-\frac{6f_\pi m_\pi^2}{\pi^2} \int_0^1 dv (1-v)\int^1_0 d\alpha_g
\int_0^{\alpha_g} d\beta
\int_0^{\beta} d\alpha  \frac{\Gamma( 1)}{(-Q^2)^{ 1}}\mid_{u=1-(1-v)\alpha_g}\nonumber\\
&&\left[V_{\parallel}+V_{\perp}+(1-2v)(A_{\parallel}+A_{\perp})
\right](\alpha,\beta,1-\alpha-\beta) \, ,
\end{eqnarray}
\begin{eqnarray}
2\sqrt{6}\Pi_\Sigma&=&-\frac{f_\pi}{3\pi^2}
 \int_0^1 du u\phi_{\pi}(u)\frac{\Gamma(\epsilon)}{(-Q^2)^\epsilon}+\frac{f_\pi m_\pi^2}{4\pi^2}
 \int_0^1 du u A(u)\frac{\Gamma(1)}{(-Q^2)^{1}}
\nonumber \\
&&-\frac{4  m_s \langle \bar{s}s\rangle f_\pi}{3}
 \int_0^1 du u\phi_{\pi}(u)\frac{\Gamma(2)}{(-Q^2)^{2}}  \nonumber \\
 &&+  \frac{2m_s \langle g_s \sigma G\bar{s}s\rangle f_\pi}{9}
 \int_0^1 du u\phi_{\pi}(u)\frac{\Gamma(3)}{(-Q^2)^{3}}\nonumber \\
 &&+ \frac{8 [\langle \bar{q}q\rangle-\langle \bar{s}s\rangle] f_\pi m_\pi^2}{9[m_u+m_d]}
 \int_0^1 du u\phi_{\sigma}(u)\frac{\Gamma(2)}{(-Q^2)^{2}}\nonumber \\
 &&- \frac{2 [\langle \bar{q}g_s \sigma Gq\rangle-\langle \bar{s}g_s \sigma Gs\rangle] f_\pi m_\pi^2}{9[m_u+m_d]}
 \int_0^1 du u\phi_{\sigma}(u)\frac{\Gamma(3)}{(-Q^2)^{3}}\nonumber \\
 &&+ \frac{ m_s f_\pi m_\pi^2}{3\pi^2[m_u+m_d]}
 \int_0^1 du u\phi_{\sigma}(u)\frac{\Gamma(1)}{(-Q^2)^{1}}\nonumber \\
&&-\frac{2f_\pi}{9}\langle\frac{ \alpha_sGG}{\pi}\rangle
 \int_0^1 du u\phi_{\pi}(u)\frac{\Gamma(2)}{(-Q^2)^{2}}\nonumber \\
&& +\frac{f_{3\pi}[8\langle \bar{q}q\rangle+11\langle \bar{s}s\rangle]}{12}\int_0^1 dvv \int_0^1 d\alpha_g \int_0^{1-\alpha_g} d\alpha_u\nonumber \\
&&   \frac{\Gamma(2)}{(-Q^2)^{2}}\mid_{u=\alpha_u+v\alpha_g} \phi_{3\pi}(\alpha_u,\alpha_g,1-\alpha_u-\alpha_g)\nonumber \\
&&+\frac{f_\pi m_\pi^2}{4\pi^2} \int_0^1 dv \int_0^1 d\alpha_g \int_0^{1-\alpha_g} d\alpha_u u \frac{\Gamma(1)}{(-Q^2)^{1}}\mid_{u=\alpha_u+v\alpha_g}\nonumber \\
&& \left[5(1-3v)V_{\perp}+4(1-2v)A_{\parallel}+4(1-v)A_{\perp} \right](\alpha_u,\alpha_g,1-\alpha_u-\alpha_g) \nonumber \\
&&+\frac{2f_\pi m_\pi^2}{\pi^2} \int_0^1 dv \int^1_0 d\alpha_g
\int_0^{1-\alpha_g} d\alpha_u
\int_0^{\alpha_u} d\alpha  \frac{\Gamma( 1)}{(-Q^2)^{ 1}}\mid_{u=\alpha_u+v\alpha_g} \nonumber\\
&&\left[V_{\parallel}+V_{\perp}+(1-2v)(A_{\parallel}+A_{\perp})
\right](\alpha,\alpha_g,1-\alpha-\alpha_g) \nonumber\\
&&-\frac{2f_\pi m_\pi^2}{\pi^2} \int_0^1 dv (1-v)\int^1_0 d\alpha_g
\int_0^{\alpha_g} d\beta
\int_0^{\beta} d\alpha  \frac{\Gamma( 1)}{(-Q^2)^{ 1}}\mid_{u=1-(1-v)\alpha_g}\nonumber\\
&&\left[V_{\parallel}+V_{\perp}+(1-2v)(A_{\parallel}+A_{\perp})
\right](\alpha,\beta,1-\alpha-\beta) \, ,
\end{eqnarray}
where $Q_\mu=q_\mu+up_\mu$ and
$Q^2=(1-u)q^2+u(p+q)^2-u(1-u)m_\pi^2$. The $\epsilon$ is a small
positive quantity, after taking the double Borel transform, we can
take the limit $\epsilon\rightarrow 0$.

 The light-cone
distribution amplitudes $\phi_{\pi}(u)$, $\phi_{\sigma}(u)$, $A(u)$,
$\phi_{3\pi}(\alpha_i)$, $A_\perp(\alpha_i)$,
$A_\parallel(\alpha_i)$, $V_\perp(\alpha_i)$ and
$V_\parallel(\alpha_i)$ of the $\pi$ meson are presented in the
appendix \cite{PSLC1,PSLC2,PSLC3,PSLC4}, the nonperturbative
parameters in the light-cone distribution amplitudes are scale
dependent, in this article, the energy scale is taken to be
$\mu=1\,\rm{GeV}$.

Taking double Borel transform  with respect to the variables
$Q_1^2=-q^2$ and $Q_2^2=-(p+q)^2$ respectively (i.e.
$\frac{\Gamma[n]}{\left[u(1-u)m_\pi^2+(1-u)Q_1^2+uQ_2^2\right]^n}
 \rightarrow \frac{M^{2(2-n)}}{M_1^2 M_2^2} e^{-\frac{u(1-u)m_\pi^2}{M^2}}
\delta(u-u_0)$, $M^2= \frac{M^2_1M^2_2}{M^2_1+M^2_2}$ and $u_0=
\frac{M_1^2}{M_1^2+M_2^2}$),  then subtract the contributions from
the high resonances and continuum states by introducing  the
threshold parameter $s_0$ (i.e. $ M^{2n}\rightarrow
\frac{1}{\Gamma[n]}\int_0^{s_0} ds s^{n-1}e^{-\frac{s}{M^2}}$),
finally we obtain two  sum rules  for the strong coupling constants
$g_{\Lambda}$ and $g_{\Sigma}$ respectively,
\begin{eqnarray}
g_{\Lambda}&=&\frac{1}{\lambda_\Lambda \lambda_{\Sigma^*}}
\exp\left\{\frac{M_{\Sigma^*}^2}{M_1^2}+\frac{M_\Lambda^2}{M_2^2}-\frac{u_0(1-u_0)m_\pi^2}{M^2}\right\}
\left\{\frac{u_0}{2\pi^2} M^4E_1(x)
 f_\pi\phi_{\pi}(u_0) \right.
\nonumber \\
&&-\frac{3u_0}{8\pi^2}M^2E_0(x)f_\pi
 m_\pi^2 A(u_0)+\frac{u_0}{3 }\langle\frac{ \alpha_sGG}{\pi}\rangle f_\pi
  \phi_{\pi}(u_0)\nonumber \\
&&+2u_0 m_s \langle \bar{s}s\rangle f_\pi \phi_\pi(u_0)-\frac{u_0
m_s\langle \bar{s}g_s \sigma Gs\rangle f_\pi\phi_\pi(u_0)}{3M^2}
\nonumber \\
&&+\frac{4u_0 [\langle \bar{q}q\rangle-\langle \bar{s}s\rangle]f_\pi
m_\pi^2 \phi_\sigma(u_0)}{9[m_u+m_d]}
+\frac{u_0 m_s M^2 E_0(x) f_\pi m_\pi^2 \phi_\sigma (u_0)}{6\pi^2[m_u+m_d]} \nonumber\\
&&-\frac{u_0[\langle \bar{q}g_s \sigma Gq\rangle-\langle \bar{s}g_s
\sigma Gs\rangle]f_\pi m_\pi^2\phi_\sigma(u_0)}
{9M^2[m_u+m_d]}\nonumber\\
 && -\frac{4\langle \bar{q}q\rangle+\langle \bar{s}s\rangle}{24}
f_{3\pi} \int_0^{u_0} d\alpha_u \int_{u_0-\alpha_u}^{1-\alpha_u}
d\alpha_g \frac{u_0-\alpha_u}{\alpha_g^2}
\phi_{3\pi}(\alpha_u,\alpha_g,1-\alpha_u-\alpha_g)\nonumber \\
&&-\frac{u_0}{8\pi^2}M^2E_0(x)f_\pi m_\pi^2 \int_0^{u_0} d\alpha_u
\int_{u_0-\alpha_u}^{1-\alpha_u} d\alpha_g \frac{1}{\alpha_g}
 \nonumber\\
&&\left[12(1-2\frac{u_0-\alpha_u}{\alpha_g})A_{\parallel}+12(1-\frac{u_0-\alpha_u}{\alpha_g})A_{\perp}\right.\nonumber\\
&&\left.-13(1-3\frac{u_0-\alpha_u}{\alpha_g})V_{\perp}\right](\alpha_u,\alpha_g,1-\alpha_u-\alpha_g) \nonumber \\
&&-\frac{3}{\pi^2}  M^2E_0(x)f_\pi m_\pi^2 \left[\int_0^{1-u_0}
d\alpha_g \int^{u_0}_{u_0-\alpha_g} d\alpha_u \int_0^{\alpha_u}
d\alpha \right.\nonumber\\
&&\left.+\int^1_{1-u_0} d\alpha_g \int^{1-\alpha_g}_{u_0-\alpha_g}
d\alpha_u
\int_0^{\alpha_u} d\alpha\right]\frac{1}{\alpha_g} \nonumber\\
&&\left[V_{\parallel}+V_{\perp}+(1-2\frac{u_0-\alpha_u}{\alpha_g})(A_{\parallel}+A_{\perp})
\right](\alpha,\alpha_g,1-\alpha-\alpha_g) \nonumber\\
&&+\frac{3}{\pi^2}  M^2E_0(x)f_\pi m_\pi^2 (1-u_0)\int_{1-u_0}^1
d\alpha_g\frac{1}{\alpha_g^2} \int_0^{\alpha_g}d\beta
\int_0^{1-\beta}d\alpha\nonumber\\
&&\left.\left[V_{\parallel}+V_{\perp}-(1-2\frac{1-u_0}{\alpha_g})(A_{\parallel}+A_{\perp})
\right](\alpha,\beta,1-\alpha-\beta)\right\} \, ,
\end{eqnarray}
\begin{eqnarray}
g_{\Sigma}&=&\frac{1}{\lambda_\Sigma \lambda_{\Sigma^*}}
\exp\left\{\frac{M_{\Sigma^*}^2}{M_1^2}+\frac{M_\Sigma^2}{M_2^2}-\frac{u_0(1-u_0)m_\pi^2}{M^2}\right\}
\left\{\frac{u_0}{2\pi^2} M^4E_1(x)
 f_\pi\phi_{\pi}(u_0) \right.
\nonumber \\
&&-\frac{3u_0}{8\pi^2}M^2E_0(x)f_\pi
 m_\pi^2 A(u_0)+\frac{u_0}{3 }\langle\frac{ \alpha_sGG}{\pi}\rangle f_\pi
  \phi_{\pi}(u_0)\nonumber \\
&&+2u_0 m_s \langle \bar{s}s\rangle f_\pi \phi_\pi(u_0)-\frac{u_0
m_s\langle \bar{s}g_s \sigma Gs\rangle f_\pi\phi_\pi(u_0)}{3M^2}
\nonumber \\
&&-\frac{4u_0 [\langle \bar{q}q\rangle-\langle \bar{s}s\rangle]f_\pi
m_\pi^2 \phi_\sigma(u_0)}
{3[m_u+m_d]}-\frac{u_0 m_s M^2 E_0(x) f_\pi m_\pi^2 \phi_\sigma (u_0)}{2\pi^2[m_u+m_d]} \nonumber\\
&&+\frac{u_0[\langle \bar{q}g_s \sigma Gq\rangle-\langle \bar{s}g_s
\sigma Gs\rangle]f_\pi m_\pi^2\phi_\sigma(u_0)}
{3M^2[m_u+m_d]}\nonumber\\
 && -\frac{8\langle \bar{q}q\rangle+11\langle \bar{s}s\rangle}{8}
f_{3\pi} \int_0^{u_0} d\alpha_u \int_{u_0-\alpha_u}^{1-\alpha_u}
d\alpha_g \frac{u_0-\alpha_u}{\alpha_g^2}
\phi_{3\pi}(\alpha_u,\alpha_g,1-\alpha_u-\alpha_g)\nonumber \\
&&-\frac{3u_0}{8\pi^2}M^2E_0(x)f_\pi m_\pi^2 \int_0^{u_0} d\alpha_u
\int_{u_0-\alpha_u}^{1-\alpha_u} d\alpha_g \frac{1}{\alpha_g}
 \nonumber\\
&&\left[4(1-2\frac{u_0-\alpha_u}{\alpha_g})A_{\parallel}+4(1-\frac{u_0-\alpha_u}{\alpha_g})A_{\perp}\right.\nonumber\\
&&\left.+5(1-3\frac{u_0-\alpha_u}{\alpha_g})V_{\perp}\right](\alpha_u,\alpha_g,1-\alpha_u-\alpha_g) \nonumber \\
&&-\frac{3}{\pi^2}  M^2E_0(x)f_\pi m_\pi^2 \left[\int_0^{1-u_0}
d\alpha_g \int^{u_0}_{u_0-\alpha_g} d\alpha_u \int_0^{\alpha_u}
d\alpha \right.\nonumber\\
&&\left.+\int^1_{1-u_0} d\alpha_g \int^{1-\alpha_g}_{u_0-\alpha_g}
d\alpha_u
\int_0^{\alpha_u} d\alpha\right]\frac{1}{\alpha_g} \nonumber\\
&&\left[V_{\parallel}+V_{\perp}+(1-2\frac{u_0-\alpha_u}{\alpha_g})(A_{\parallel}+A_{\perp})
\right](\alpha,\alpha_g,1-\alpha-\alpha_g) \nonumber\\
&&+\frac{3}{\pi^2}  M^2E_0(x)f_\pi m_\pi^2 (1-u_0)\int_{1-u_0}^1
d\alpha_g\frac{1}{\alpha_g^2} \int_0^{\alpha_g}d\beta
\int_0^{1-\beta}d\alpha\nonumber\\
&&\left.\left[V_{\parallel}+V_{\perp}-(1-2\frac{1-u_0}{\alpha_g})(A_{\parallel}+A_{\perp})
\right](\alpha,\beta,1-\alpha-\beta)\right\} \, ,
\end{eqnarray}
where
\begin{eqnarray}
E_n(x)&=&1-(1+x+\frac{x^2}{2!}+\cdots+\frac{x^n}{n!})e^{-x} \, , \nonumber\\
x&=&\frac{s_0}{M^2} \, .\nonumber
\end{eqnarray}

\section{Numerical result and discussion}

The input parameters are taken as
$m_u=m_d=(0.0056\pm0.0016)\,\rm{GeV}$, $f_\pi=0.130\,\rm{GeV}$,
$m_{\pi} =0.138\,\rm{GeV}$, $\lambda_3=0.0$ (which appears in the
coefficient of the three-particle light-cone distribution amplitude
$\phi_{3\pi}(\alpha_i)$, one can consult Ref.\cite{PSLC4} for the
definition), $f_{3\pi}=(0.45\pm0.15)\times 10^{-2}\,\rm{GeV}^2$,
$\omega_3=-1.5\pm0.7$, $\omega_4=0.2\pm0.1$, $a_2=0.25\pm 0.15$,
$a_1=0.0 $, $\eta_4=10.0\pm3.0$ \cite{PSLC1,PSLC2,PSLC3,PSLC4},
$\langle \bar{q}q \rangle=-(0.24\pm 0.01\, \rm{GeV})^3$, $\langle
\bar{s}s \rangle=(0.8\pm 0.2 )\langle \bar{q}q \rangle$, $\langle
\bar{q}g_s\sigma Gq \rangle=m_0^2\langle \bar{q}q \rangle$, $\langle
\bar{s}g_s\sigma Gs \rangle=m_0^2\langle \bar{s}s \rangle$,
$m_0^2=(0.8 \pm 0.2)\,\rm{GeV}^2$, $\langle \frac{\alpha
GG}{\pi}\rangle=(0.33\,\rm{GeV})^4 $ \cite{SVZ79,PRT85},
$M_{\Sigma^*}=1.3828\,\rm{GeV}$, $M_{\Sigma}=1.1926\,\rm{GeV}$,
$M_{\Lambda}=1.1157\,\rm{GeV}$ \cite{PDG},
$\lambda_{\Lambda}=(2.7\pm 0.2)\times 10^{-2}\,\rm{GeV}^3$,
$\lambda_{\Sigma}=(2.8\pm 0.2)\times 10^{-2}\,\rm{GeV}^3$ and
$\lambda_{\Sigma^*}=(3.7\pm0.2)\times 10^{-2}\,\rm{GeV}^3$
\cite{Ioffe1,Ioffe2,Ioffe3,Ioffe4}.

In this article, we neglect  the  perturbative $\mathcal
{O}(\alpha_s)$ corrections to the strong coupling constants
$g_{\Sigma^* \Lambda/\Sigma \pi}$, and take the values of the pole
residues $\lambda_{\Sigma}$, $\lambda_{\Lambda}$ and
$\lambda_{\Sigma^*}$ without perturbative $\mathcal {O}(\alpha_s)$
corrections  for consistency.

The threshold parameter $s_0$ is chosen to be
$s_0=(3.6\pm0.1)\,\rm{GeV}^2$ to avoid possible contamination from
the contributions of the high resonance states, and it is large
enough to take into account  the contribution of the decuplet baryon
$\Sigma^*$. Although the $P_{13}$ state $\Sigma (1840)$ and the
$P_{11}$ states $\Sigma(1770)$, $\Sigma(1880)$ are below the
threshold, they have no contaminations due to the mismatch of the
isospin and spin \cite{PDG}.

The Borel parameters are chosen as
$\frac{M_{\Sigma^*}^2}{M_1^2}=\frac{M_{\Lambda/\Sigma}^2}{M_2^2}$
and
 $M^2=\frac{M_1^2M^2_{\Lambda/\Sigma}}{M_{\Sigma^*}^2+M^2_{\Lambda/\Sigma}}=(2.2-3.2)\,\rm{GeV}^2$,
 in those regions,
the value of the strong coupling constants $g_{\Lambda}$ and
$g_{\Sigma}$ are rather stable with  variation of the Borel
parameter $M^2$.

The theoretical values of the $a_2$  vary in a large range
($a_2=0.10\sim 0.40$) at the energy scale $\mu=1\, \rm{GeV}$
\cite{PSLC4},  we can take smaller uncertainty, say $30\%$ (i.e.
$a_2=0.25\pm 0.08$), which is the typical uncertainty in the QCD sum
rules.  The value obtained by Ball, Braun and Lenz with the QCD sum
rules is $a_2=0.28 \pm 0.08$ \cite{PSLC4}, which has the typical
uncertainty. In this article, we present the results with two sets
of  parameters,  the parameters  characterized by  $a_2=0.25\pm
0.08$ and $a_2=0.28 \pm 0.08$ are denoted as $\rm{P\,I}$ and
$\rm{P\,II}$ respectively, because other parameters have the same
values.

In calculation, we observe  the main uncertainties come from the two
parameters $a_2$ and $\eta_4$, the uncertainty originates from the
parameter $\omega_4$ is also considerable, which are shown in
Figs.1-3.

The dominant contributions come from the two-particle light-cone
distribution amplitudes $\phi_\pi(u)$ and $A(u)$; the contributions
from the terms involving the three-particle (quark-antiquark-gluon)
light-cone distribution amplitudes  are of minor importance, about
$7\%$ and $12\%$ of the contribution from the term
$\frac{u_0}{2\pi^2}M^4E_1(x)f_\pi
 \phi_{\pi}(u_0)$ for the $g_\Lambda$ and $g_\Sigma$ respectively.

 The shapes of the   light-cone distribution amplitudes  $\phi_\pi(u)$
 and $A(u)$ have significant impacts on the values of the  $g_\Lambda$ and
 $g_\Sigma$, because only the values of the special point $u=u_0$
 are  involved.  This case is in contrast to  the light-cone QCD sum rules for the hadronic form-factors,
  where the momentum fraction $u$ is integrated out,
  dependence on the shapes is  mild.  For example, the $\phi_\pi(u)$
has been  analyzed with the light-cone QCD sum rules and (non-local
condensates) QCD sum rules confronting with the high precision CLEO
data on the $  \gamma \gamma^* \to \pi^0$ transition form-factor
\cite{Schmedding1999-24,Bakulev2001-24,Bakulev2002-24,Bakulev2003-24,Bakulev2006-24,Bakulev2006-24-2,Braun00,Bijnens02},
where the $\phi_\pi(u)$ is expanded in terms of the Gegenbauer
polynomials $ C_n^{\frac{3}{2}}(2u-1)$, truncations  at the order $
n=2$ or  $ n=4$  both  lead to satisfactory results.

The strong coupling constant $g$ can serve as an excellent subject
for determining  the shapes of  the light-cone distributions
amplitudes $\phi_\pi(u)$
 and $A(u)$,   perturbative $\mathcal {O}(\alpha_s)$ corrections
should be taken into account  before confronting with the
experimental data. To my knowledge, only the leading order
contributions to the strong
 coupling constants  $g_{ NNV}$ and $g_{NNP}$ have been calculated with the
 light-cone QCD sum rules \cite{Baryon1,Baryon2,Baryon3,Baryon4}, where the $N$, $V$ and $P$ denote
 the octet baryons, the vector mesons and the pseudoscalar mesons, respectively.

\begin{figure}
\centering
 \includegraphics[totalheight=6cm,width=7cm]{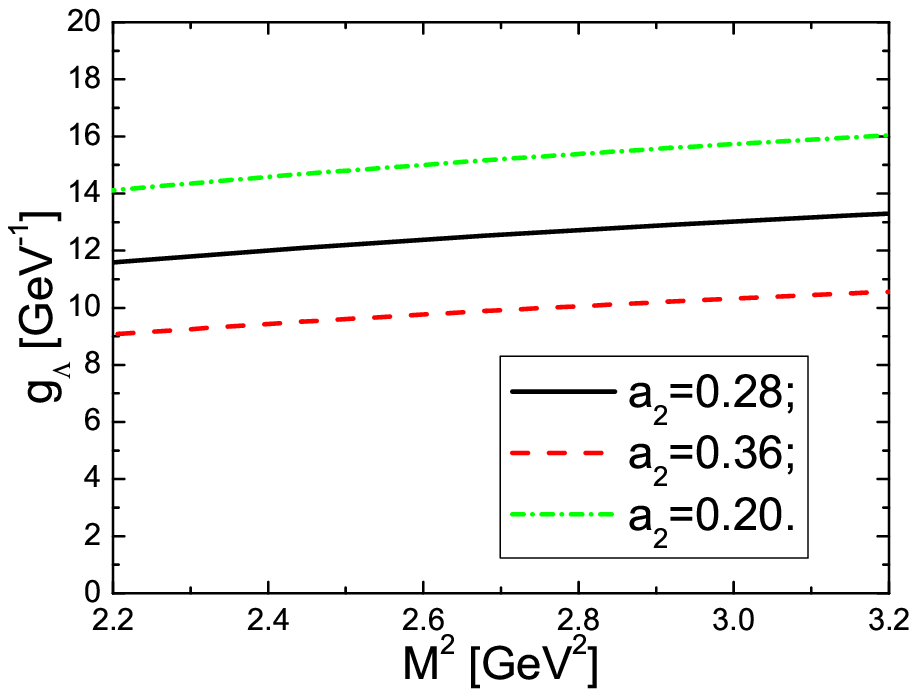}
   \includegraphics[totalheight=6cm,width=7cm]{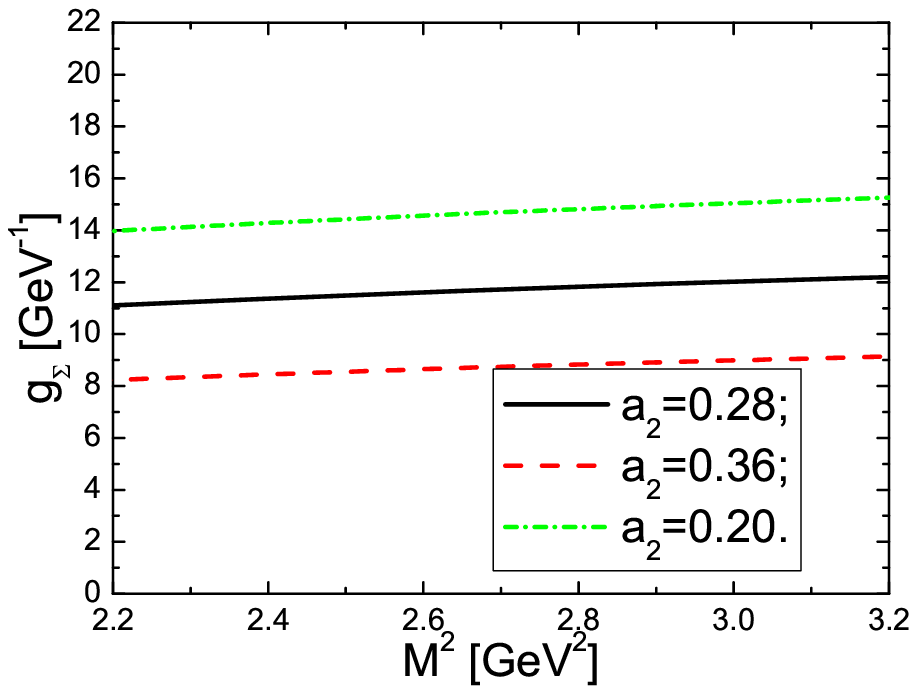}
           \caption{ The strong coupling constants $g_{\Lambda}$ and $g_{\Sigma}$ with variation of the
           Borel parameter $M^2$ and the coefficient $a_2$. }
\end{figure}

\begin{figure}
\centering
 \includegraphics[totalheight=6cm,width=7cm]{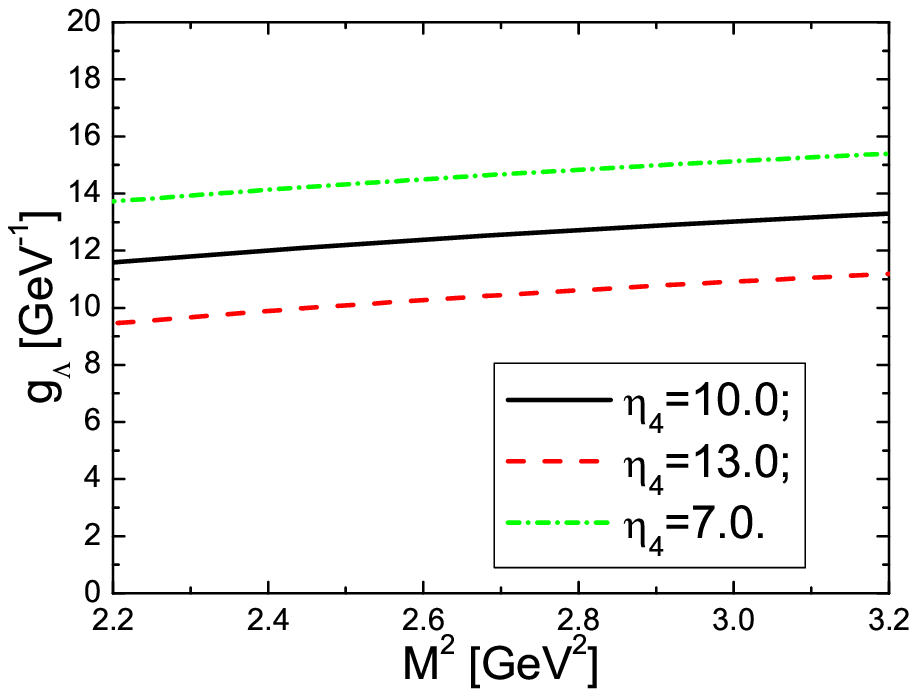}
   \includegraphics[totalheight=6cm,width=7cm]{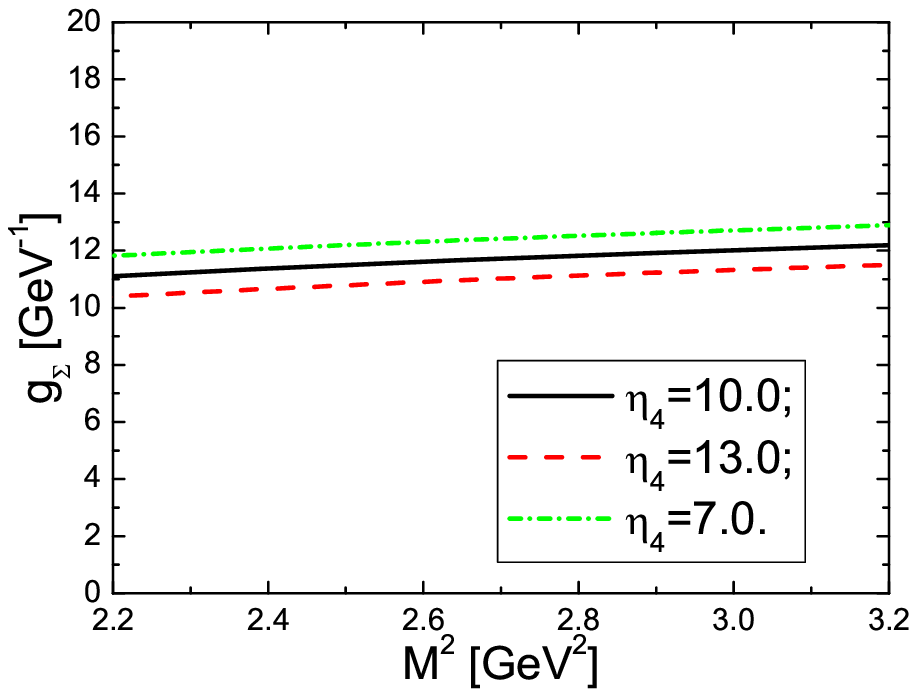}
           \caption{ The strong coupling constants $g_{\Lambda}$ and $g_{\Sigma}$ with variation of the
           Borel parameter $M^2$ and the nonperturbative parameter  $\eta_4$. }
\end{figure}

\begin{figure}
\centering
 \includegraphics[totalheight=6cm,width=7cm]{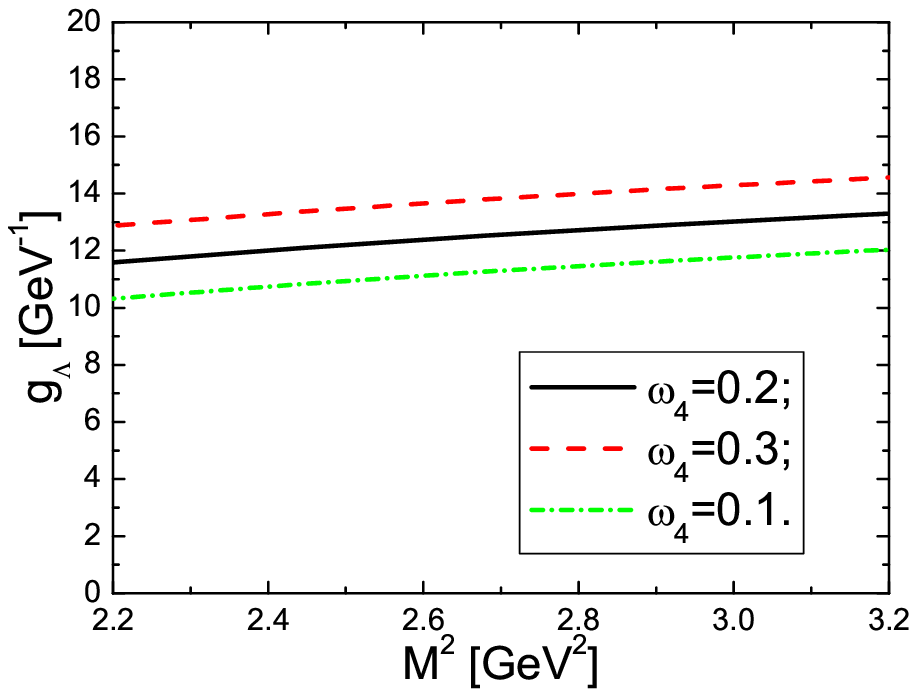}
   \includegraphics[totalheight=6cm,width=7cm]{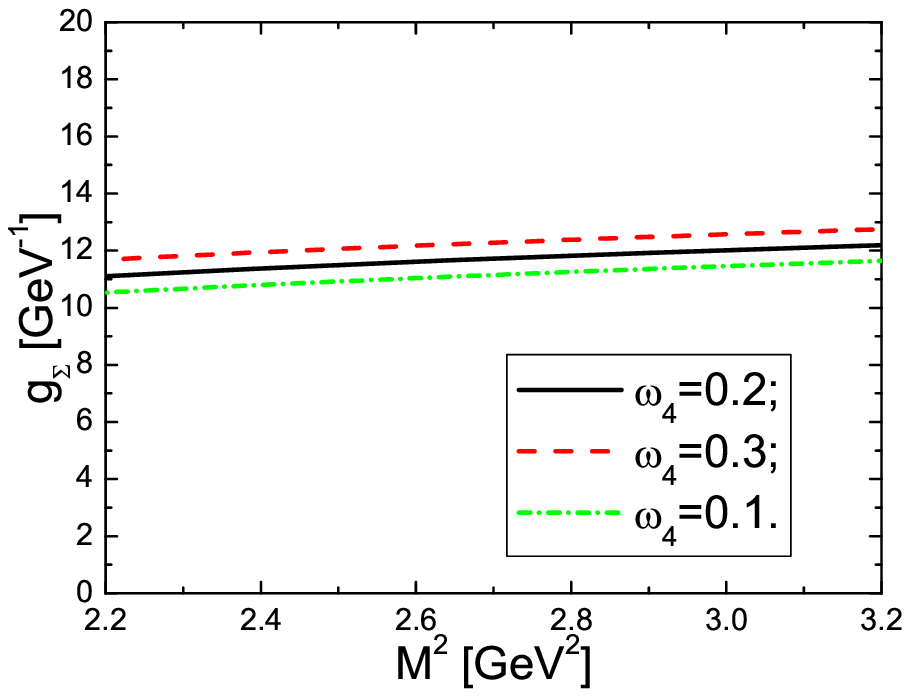}
           \caption{ The strong coupling constants $g_{\Lambda}$ and $g_{\Sigma}$ with variation of the
           Borel parameter $M^2$ and the nonperturbative parameter  $\omega_4$. }
\end{figure}

Taking into account all the uncertainties, finally we obtain the
numerical results for the strong coupling constants $g_\Lambda$ and
$g_\Sigma$, which are shown in  Figs.4-5,
\begin{eqnarray}
  g_{\Lambda} &=&(13.6\pm 4.6)\,\rm{GeV}^{-1} \, \, , \nonumber \\
  g_{\Sigma} &=&(12.9\pm 4.2)\,\rm{GeV}^{-1} \, \, ,\nonumber \\
  g_{N } &=&(13.5\pm 5.4)\,\rm{GeV}^{-1}\, \, ,
\end{eqnarray}
and
\begin{eqnarray}
  g_{\Lambda} &=&(12.6\pm 4.7)\,\rm{GeV}^{-1} \, \, , \nonumber \\
  g_{\Sigma} &=&(11.8\pm 4.1)\,\rm{GeV}^{-1} \, \, ,\nonumber \\
  g_{N } &=&(12.5\pm 5.5)\,\rm{GeV}^{-1}\, \, ,
\end{eqnarray}
for the parameters $\rm{P\,I}$ and $\rm{P\,II}$ respectively, here
we also present the value of the strong coupling constant $g_{\Delta
p \pi}$ with the light-cone QCD sum rules \cite{Wang0707}. We
calculate uncertainties $\delta$  with the formula
$\delta=\sqrt{\sum_i\left(\frac{\partial f}{\partial x_i}\right)^2
(x_i-\bar{x}_i)^2}$, where the $f$ denote the strong coupling
constants $g_\Lambda$, $g_\Sigma$ and $g_{N }$, the $x_i$ denote the
input parameters $m_u$, $m_d$, $a_2$, $f_{3\pi}$, $\cdots$. The
average values are
\begin{eqnarray}
g&=&13.3\pm 4.7 \, \rm{GeV}^{-1}\, , \nonumber\\
\mathcal{C}&=&1.7\pm 0.6 \, ,
\end{eqnarray}
and
\begin{eqnarray}
g&=&12.3\pm 4.7 \, \rm{GeV}^{-1}\, , \nonumber\\
\mathcal{C}&=&1.6\pm 0.6 \, .
\end{eqnarray}
for the parameters $\rm{P\,I}$ and $\rm{P\,II}$ respectively.

 The uncertainties are rather large, larger than
$30\%$, and no definitive  conclusion can be drawn for the $SU(3)$
breaking effects.  If we take the central values as the input
parameters, the $SU(3)$ breaking effects are rather small, less than
$6\%$. The uncertainties may result in larger $SU(3)$ breaking
effects, furthermore, we have neglected the perturbative $\mathcal
{O}(\alpha_s)$ corrections, which may also contribute to the $SU(3)$
breaking effects.

\begin{figure}
\centering
 \includegraphics[totalheight=6cm,width=7cm]{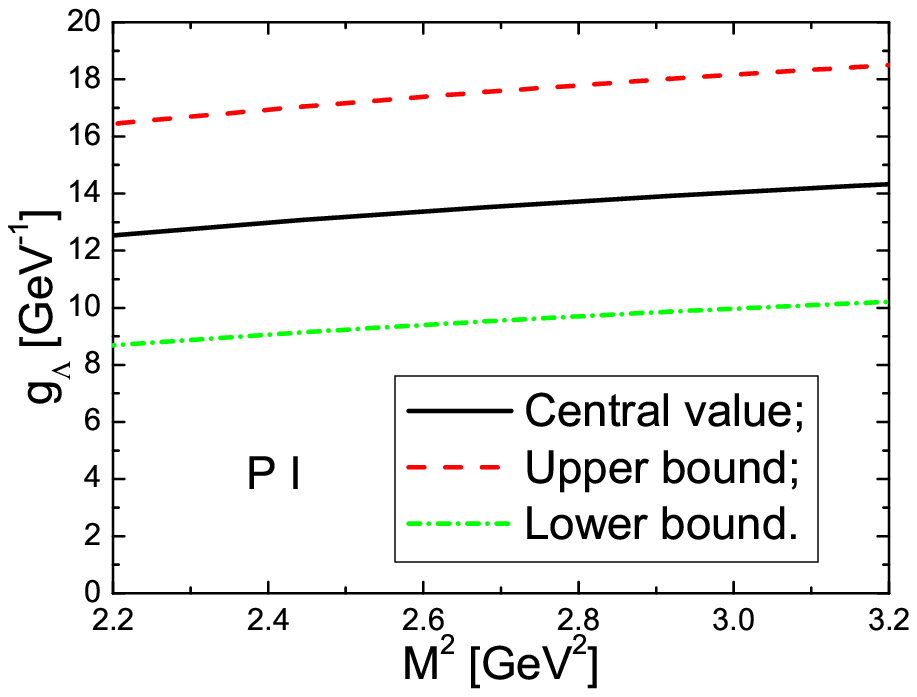}
   \includegraphics[totalheight=6cm,width=7cm]{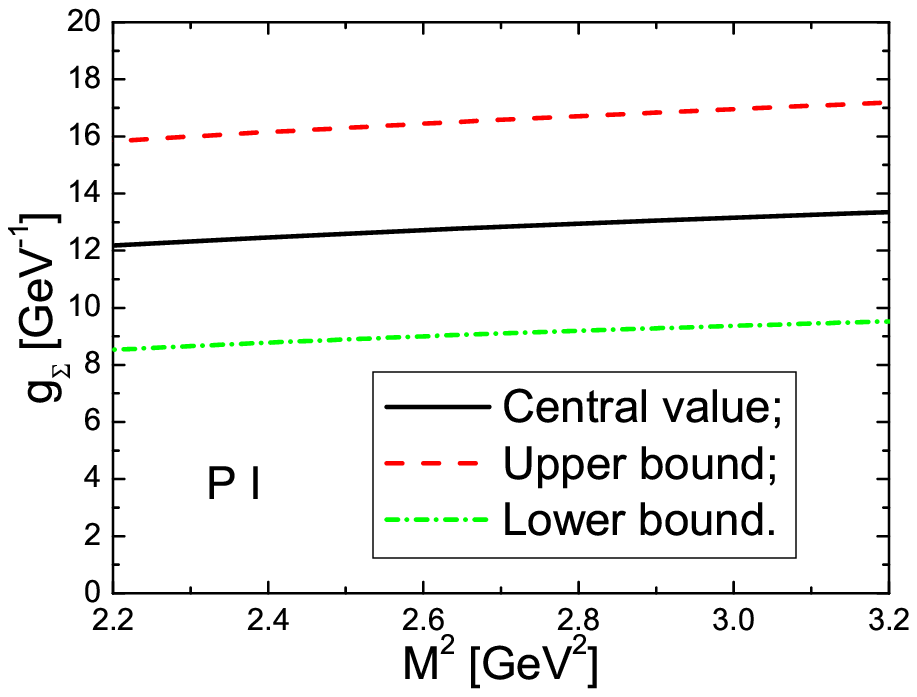}
           \caption{ The strong coupling constants $g_{\Lambda}$ and $g_{\Sigma}$ with variation of the
           Borel parameter $M^2$ for the parameters $\rm{P\,I}$. The uncertainties $\delta$  are calculated
           with the formula
$\delta=\sqrt{\sum_i\left(\frac{\partial f}{\partial x_i}\right)^2
(x_i-\bar{x}_i)^2}$, where the $f$ denote the strong coupling
constants $g_\Lambda$ and $g_\Sigma$, the $x_i$ denote the input
parameters $m_u$,  $a_2$, $f_{3\pi}$, $\cdots$. }
\end{figure}

\begin{figure}
\centering
 \includegraphics[totalheight=6cm,width=7cm]{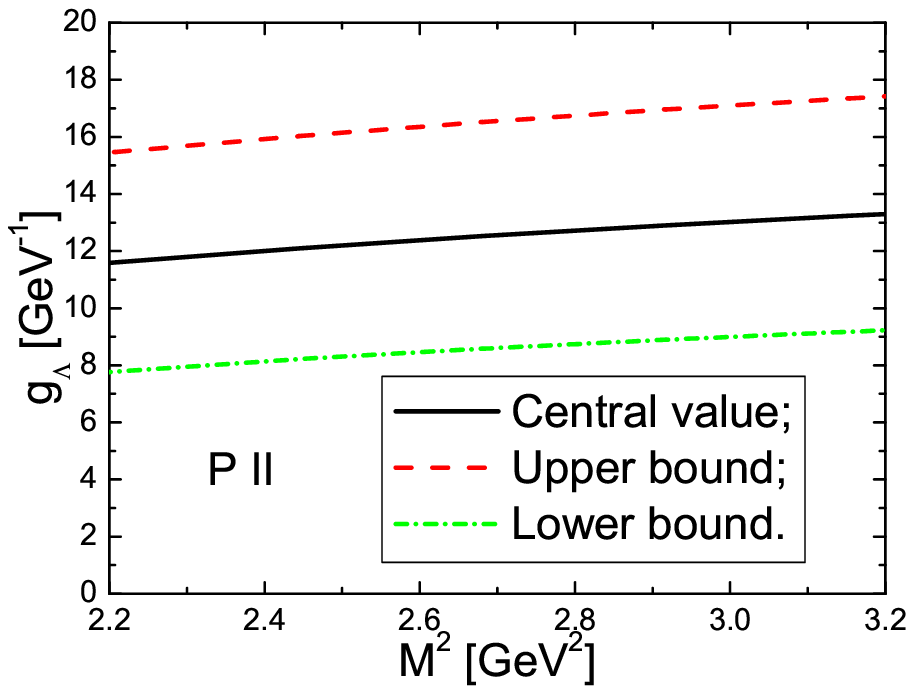}
   \includegraphics[totalheight=6cm,width=7cm]{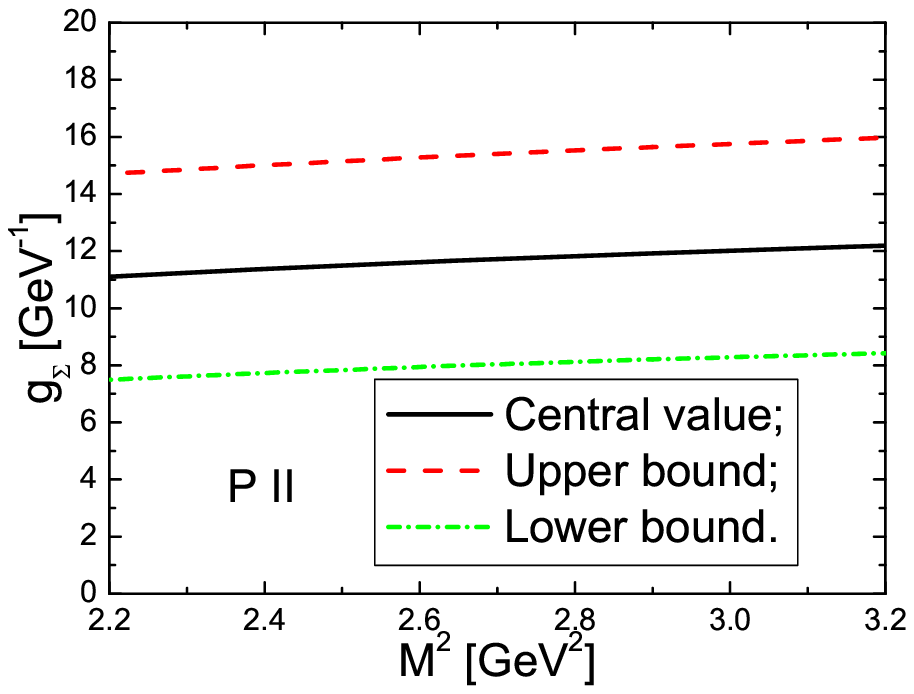}
           \caption{ The strong coupling constants $g_{\Lambda}$ and $g_{\Sigma}$ with variation of the
           Borel parameter $M^2$ for the parameters $\rm{P\,II}$. The uncertainties $\delta$  are calculated
           with the formula
$\delta=\sqrt{\sum_i\left(\frac{\partial f}{\partial x_i}\right)^2
(x_i-\bar{x}_i)^2}$, where the $f$ denote the strong coupling
constants $g_\Lambda$ and $g_\Sigma$, the $x_i$ denote the input
parameters $m_u$,  $a_2$, $f_{3\pi}$, $\cdots$.}
\end{figure}

The strong coupling constants $g_{\Sigma^* \Lambda/\Sigma\pi}$ have
the following relation with the decay widths $\Gamma_{\Sigma^* \to
\Lambda/\Sigma\pi}$,
\begin{eqnarray}
\Gamma_{\Sigma^* \to \Lambda/\Sigma \pi}&=&\frac{g^2_{\Sigma^*
\Lambda/\Sigma \pi} p_{cm}}{32\pi M_{\Sigma^*}^2} \sum_{ss'} \mid
\overline{U}(p',s) p^\mu_\pi U_\mu(p'',s') \mid^2 \, ,
\nonumber \\
p_{cm}&=&\frac{\sqrt{[M_{\Sigma^*}^2-(M_{\Lambda/\Sigma}+m_\pi)^2][M_{\Sigma^*}^2-(M_{\Lambda/\Sigma}-m_\pi)^2]}}
{2M_{\Sigma^*}} \, .
\end{eqnarray}
 If we take the experimental data as the input
parameters, $\Gamma_{\Sigma^* \to \Sigma\pi}=4.19\,\rm{MeV}$,
$\Gamma_{\Sigma^* \to \Lambda\pi}=31.15\,\rm{MeV}$ and
$\Gamma_{\Delta \to p\pi}=118.0\,\rm{MeV}$ \cite{PDG}, we can obtain
the values $g_{\Sigma}\approx 17.4 \,\rm{GeV}^{-1}$,
$g_{\Lambda}\approx 12.8 \,\rm{GeV}^{-1}$ and $g_{ N }\approx  15.6
\,\rm{GeV}^{-1}$.  The average value is about $g\approx 15.3\,
\rm{GeV}^{-1}$, and the $SU(3)$ breaking effects are  about
$(12-18)\%$. The values $\Gamma_{\Sigma^* \to
\Sigma\pi}=4.19\,\rm{MeV}$ and $\Gamma_{\Sigma^* \to
\Lambda\pi}=31.15\,\rm{MeV}$ are estimated (not fitted or averaged)
by the Particle Data Group  \cite{PDG}; more accurate data may
result in smaller $SU(3)$ breaking effects.

In the region $M^2=(2.2-3.2)\,\rm{GeV}^2$,
$\frac{\alpha_s(M)}{\pi}\sim 0.10-0.12$ \cite{AlphaS}. If the
radiative $\mathcal {O}(\alpha_s)$ corrections to the leading
perturbative terms are companied  with large numerical factors, just
like in the case of the QCD sum rules for the mass of the proton
\cite{Ioffe2005}, $ 1+(\frac{53}{12}+\gamma_E)
\frac{\alpha_s(M)}{\pi} \sim 1+(0.53-0.62) $, the contributions of
the order $\mathcal {O}(\alpha_s)$ are large.  Furthermore, the pole
residues $\lambda_\Lambda$, $\lambda_\Sigma$ and $\lambda_{\Sigma^*
}$ also receive contributions from the perturbative $\mathcal
{O}(\alpha_s)$ corrections, if  they are taken into account
properly, we can improve the value of the strong coupling constant
$g$.

\section{Conclusion}

In this article, we calculate the strong coupling constant $g$ among
the decuplet baryons, the octet baryons and the pseudoscalar mesons
in the heavy baryon chiral perturbation theory with the light-cone
QCD sum rules, and study the strong decays $\Sigma^* \to \Lambda
\pi,\Sigma \pi$. The numerical value of the strong coupling constant
$g$ is consistent with our previous calculation, the central values
lead to  small $SU(3)$ breaking effects, less than $6\%$; and no
definitive  conclusion can be drawn due to the large uncertainties.
The perturbative $\mathcal {O}(\alpha_s)$ corrections may improve
the results further.

 \section*{Appendix}
 The light-cone distribution amplitudes of the $\pi$ meson are defined
 by \cite{PSLC1,PSLC2,PSLC3,PSLC4}
\begin{eqnarray}
\langle0| {\bar u} (x) \gamma_\mu \gamma_5 d(0) |\pi(p)\rangle& =& i
f_\pi p_\mu \int_0^1 du  e^{-i u p\cdot x}
\left\{\phi_\pi(u)+\frac{m_\pi^2x^2}{16}
A(u)\right\}\nonumber\\
&&+\frac{i}{2}f_\pi m_\pi^2\frac{x_\mu}{p\cdot x}
\int_0^1 du  e^{-i u p \cdot x} B(u) \, , \nonumber\\
\langle0| {\bar u} (x) i \gamma_5 d(0) |\pi(p)\rangle &=&
\frac{f_\pi m_\pi^2}{
m_u+m_d}\int_0^1 du  e^{-i u p \cdot x} \phi_p(u)  \, ,  \nonumber\\
\langle0| {\bar u} (x) \sigma_{\mu \nu} \gamma_5 d(0) |\pi(p)\rangle
&=&i(x_\mu p_\nu-x_\nu p_\mu)  \frac{f_\pi m_\pi^2}{6 (m_u+m_d)}
\int_0^1 du
e^{-i u p \cdot x} \phi_\sigma(u) \, ,  \nonumber\\
\langle0| {\bar u} (x) \sigma_{\mu \nu} \gamma_5 g_s G_{\alpha \beta
}(v x)d(0) |\pi(p)\rangle&=& f_{3 \pi}\left\{(p_\mu p_\alpha
g^\bot_{\nu
\beta}-p_\nu p_\alpha g^\bot_{\mu \beta}) -(p_\mu p_\beta g^\bot_{\nu \alpha}\right.\nonumber\\
&&\left.-p_\nu p_\beta g^\bot_{\mu \alpha})\right\} \int {\cal
D}\alpha_i \phi_{3\pi} (\alpha_i)
e^{-ip \cdot x(\alpha_u+v \alpha_g)} \, ,\nonumber\\
\langle0| {\bar u} (x) \gamma_{\mu} \gamma_5 g_s G_{\alpha
\beta}(vx)d(0) |\pi(p)\rangle&=&  f_\pi m_\pi^2p_\mu  \frac{p_\alpha
x_\beta-p_\beta x_\alpha}{p
\cdot x}\nonumber\\
&&\int{\cal D}\alpha_i A_{\parallel}(\alpha_i) e^{-ip\cdot
x(\alpha_u +v \alpha_g)}\nonumber \\
&&+ f_\pi m_\pi^2 (p_\beta g^\perp_{\alpha\mu}-p_\alpha
g^\perp_{\beta\mu})\nonumber\\
&&\int{\cal D}\alpha_i A_{\perp}(\alpha_i)
e^{-ip\cdot x(\alpha_u +v \alpha_g)} \, ,  \nonumber\\
\langle0| {\bar u} (x) \gamma_{\mu} i g_s \tilde G_{\alpha
\beta}(vx)d(0) |\pi(p)\rangle&=& f_\pi m_\pi^2 p_\mu  \frac{p_\alpha
x_\beta-p_\beta x_\alpha}{p \cdot
x}\nonumber\\
&&\int{\cal D}\alpha_i V_{\parallel}(\alpha_i) e^{-ip\cdot
x(\alpha_u +v \alpha_g)}\nonumber \\
&&+ f_\pi m_\pi^2 (p_\beta g^\perp_{\alpha\mu}-p_\alpha g^\perp_{\beta\mu})\nonumber\\
&&\int{\cal D}\alpha_i V_{\perp}(\alpha_i) e^{-ip\cdot x(\alpha_u +v
\alpha_g)} \, ,
\end{eqnarray}
where $g_{\mu\nu}^\perp=g_{\mu\nu}-\frac{p_\mu x_\nu+p_\nu x_\mu}{p
\cdot x}$, $\tilde G_{\mu \nu}= \frac{1}{2} \epsilon_{\mu\nu
\alpha\beta} G^{\alpha\beta} $ and ${\cal{D}} \alpha_i =d \alpha_u d
\alpha_d d \alpha_g \delta(1-\alpha_u -\alpha_d -\alpha_g)$.

The light-cone distribution amplitudes of the $\pi$ meson are
parameterized as \cite{PSLC1,PSLC2,PSLC3,PSLC4}
\begin{eqnarray}
\phi_\pi(u)&=&6u(1-u)
\left\{1+a_1C^{\frac{3}{2}}_1(\xi)+a_2C^{\frac{3}{2}}_2(\xi)
\right\}\, , \nonumber\\
\phi_p(u)&=&1+\left\{30\eta_3-\frac{5}{2}\rho^2\right\}C_2^{\frac{1}{2}}(\xi)\nonumber \\
&&+\left\{-3\eta_3\omega_3-\frac{27}{20}\rho^2-\frac{81}{10}\rho^2 a_2\right\}C_4^{\frac{1}{2}}(\xi)\, ,  \nonumber \\
\phi_\sigma(u)&=&6u(1-u)\left\{1
+\left[5\eta_3-\frac{1}{2}\eta_3\omega_3-\frac{7}{20}\rho^2-\frac{3}{5}\rho^2 a_2\right]C_2^{\frac{3}{2}}(\xi)\right\}\, , \nonumber \\
\phi_{3\pi}(\alpha_i) &=& 360 \alpha_u \alpha_d \alpha_g^2 \left \{1
+\lambda_3(\alpha_u-\alpha_d)+ \omega_3 \frac{1}{2} ( 7 \alpha_g
- 3) \right\} \, , \nonumber\\
V_{\parallel}(\alpha_i) &=& 120\alpha_u \alpha_d \alpha_g \left(
v_{00}+v_{10}(3\alpha_g-1)\right)\, ,
\nonumber \\
A_{\parallel}(\alpha_i) &=& 120 \alpha_u \alpha_d \alpha_g a_{10}
(\alpha_d-\alpha_u)\, ,
\nonumber\\
V_{\perp}(\alpha_i) &=& -30\alpha_g^2
\left\{h_{00}(1-\alpha_g)+h_{01}\left[\alpha_g(1-\alpha_g)-6\alpha_u
\alpha_d\right] \right.  \nonumber\\
&&\left. +h_{10}\left[
\alpha_g(1-\alpha_g)-\frac{3}{2}\left(\alpha_u^2+\alpha_d^2\right)\right]\right\}\,
, \nonumber\\
A_{\perp}(\alpha_i) &=&  30 \alpha_g^2 (\alpha_u-\alpha_d) \left\{h_{00}+h_{01}\alpha_g+\frac{1}{2}h_{10}(5\alpha_g-3)  \right\}, \nonumber\\
A(u)&=&6u(1-u)\left\{
\frac{16}{15}+\frac{24}{35}a_2+20\eta_3+\frac{20}{9}\eta_4 \right.
\nonumber \\
&&+\left[
-\frac{1}{15}+\frac{1}{16}-\frac{7}{27}\eta_3\omega_3-\frac{10}{27}\eta_4\right]C^{\frac{3}{2}}_2(\xi)
\nonumber\\
&&\left.+\left[
-\frac{11}{210}a_2-\frac{4}{135}\eta_3\omega_3\right]C^{\frac{3}{2}}_4(\xi)\right\}+\left\{
 -\frac{18}{5}a_2+21\eta_4\omega_4\right\} \nonumber\\
 && \left\{2u^3(10-15u+6u^2) \log u+2\bar{u}^3(10-15\bar{u}+6\bar{u}^2) \log \bar{u}
 \right. \nonumber\\
 &&\left. +u\bar{u}(2+13u\bar{u})\right\} \, ,\nonumber\\
 g(u)&=&1+g_2C^{\frac{1}{2}}_2(\xi)+g_4C^{\frac{1}{2}}_4(\xi)\, ,\nonumber\\
 B(u)&=&g(u)-\phi_\pi(u)\, ,
\end{eqnarray}
where
\begin{eqnarray}
h_{00}&=&v_{00}=-\frac{\eta_4}{3} \, ,\nonumber\\
a_{10}&=&\frac{21}{8}\eta_4 \omega_4-\frac{9}{20}a_2 \, ,\nonumber\\
v_{10}&=&\frac{21}{8}\eta_4 \omega_4 \, ,\nonumber\\
h_{01}&=&\frac{7}{4}\eta_4\omega_4-\frac{3}{20}a_2 \, ,\nonumber\\
h_{10}&=&\frac{7}{2}\eta_4\omega_4+\frac{3}{20}a_2 \, ,\nonumber\\
g_2&=&1+\frac{18}{7}a_2+60\eta_3+\frac{20}{3}\eta_4 \, ,\nonumber\\
g_4&=&-\frac{9}{28}a_2-6\eta_3\omega_3 \, ,
\end{eqnarray}
   $\xi=2u-1$, and $ C_2^{\frac{1}{2}}(\xi)$, $ C_4^{\frac{1}{2}}(\xi)$,
 $ C_1^{\frac{3}{2}}(\xi)$, $ C_2^{\frac{3}{2}}(\xi)$ are Gegenbauer polynomials,
  $\eta_3=\frac{f_{3\pi}}{f_\pi}\frac{m_u+m_d}{m_\pi^2}$ and  $\rho^2={(m_u+m_d)^2\over m_\pi^2}$
 \cite{PSLC1,PSLC2,PSLC3,PSLC4}.

\section*{Acknowledgements}
This  work is supported by National Natural Science Foundation,
Grant Number 10775051, and Program for New Century Excellent Talents
in University, Grant Number NCET-07-0282.

\end{document}